\documentclass{ws-ijmpd}
\usepackage[super,compress]{cite}
\usepackage{orcidlink}
\begin{document}

\markboth{Ziad Sakr}
{The Case for a Low Matter Density}

%
\catchline{}{}{}{}{}
%

\title{THE CASE FOR A LOW DARK MATTER DENSITY IN DYNAMICAL DARK ENERGY MODEL FROM LOCAL PROBES 
 }
\newcommand{\orcid}[1]{\orcidlink{#1}}
\author{ZIAD SAKR  \orcid{0000-0002-4823-3757} } 

\address{Faculty of Sciences, University Saint Joseph, \\
Beirut,Lebanon\\
ziad.sakr@net.usj.edu.lb\\
IRAP, Université de Toulouse, CNRS, CNES, UPS,\\
Toulouse, France}

%

\maketitle


\begin{abstract}
In this work we investigate, through a Bayesian study, the ability of a local low matter density $\Omega_{\rm M}$, in discrepancy with the value usually inferred from the CMB angular power spectrum, to accommodate observations from local probes without being in tension with the local values of the Hubble constant $H_0$ or the matter fluctuation $\sigma_8$ parameters. For that, we combine multiple local probes, with the criteria that they either can constrain the matter density parameter independently from the CMB constraints, or can help in doing so after making their relevant observations more model independent by relaxing their relevant calibration parameters. We assume however, either a dynamical dark energy model, or  the standard $\Lambda$CDM  model, when computing the corresponding theoretical observables. We also add the latest Baryonic acoustic oscillations (BAO) measurements from the DESI year one release to some of our MCMC runs. We found that, within $\Lambda$CDM model, for different combinations of our probes, we can accommodate a low matter density  along with the $H_0$ and $\sigma_8$ values usually obtained from local probes, providing we promote the sound drag $r_s$ component in BAO calculations to a free parameter, and that even if we combine with the Pantheon+ Supernova sample, an addition that was found in previous work to mitigate our concordance. Assuming $w_0w_a$CDM, we also found that relaxing $r_s$ allow us to accommodate $\Omega_{\rm M}$, $H_0$ and $\sigma_8$ within their local values, with still however a preference for $w_0w_a$ values far from $\Lambda$CDM. However, when including Pantheon+ Supernova sample, we found that the latter preference for high matter density pushes $\sigma_8$ to values much smaller than the ones usually obtained from local probes, mitigating by then a low matter density solution to the two common tensions. We conclude therefore that a low matter density value, helps in preserving the concordance within $\Lambda$CDM model, even with recent BAO DESI measurements or high redshift supernovae sample, while the dynamical dark energy model ultimately fails to solve all tensions at once within our low density matter hypothesis. 
\end{abstract}


\ccode{PACS numbers:}


\section{Introduction}	

The longly successful $\Lambda$CDM model, dubbed as the cosmological standard model, is more and more facing a lack of concordance between some of its cosmological parameters when the latter are constrained with different probes. Namely the Hubble constant $H_0$ and the matter fluctuation parameter $\sigma_8$. Several solutions have been proposed to alleviate these tensions without full success, especially when trying to alleviate both at the same time (see Sakr \cite{Sakr:2023hrl} (2022) and reference therein for a review of many of the proposed solutions, or, for more recent ones \cite{Jimenez:2024lmm,Brito:2024bhh,Tang:2024lmo,Mirpoorian:2024fka,Pang:2024wul,Ye:2024zpk,Zhumabek:2024tvp,DeSimone:2024lvy,Simon:2024jmu,Toda:2024uff,SolaPeracaula:2024iil,Toda:2024ncp,Yadav:2024duq,Liu:2024vlt,zabat:2024wof,HosseiniMansoori:2024pdq,Sakr:2024eee,Gomez-Valent:2024tdb,Cook:2024eff,Yeung:2024krv,Akarsu:2024qiq,Khosravi:2023rhy,Tutusaus:2023cms,Sakr:2023mao,Yarahmadi:2024lzd,Shah:2024gfu,Barua:2024gei,Sadeghnezhad:2025txo,Yarahmadi:2024afr}). Recently, there have been indications about a preference for a dynamical dark energy equation of state parameters ($w_0,w_a$) far from their  $\Lambda$CDM limit values when constrained using a combination of datasets from Cosmic Microwave Background from the Planck 2018 Mission (hereafter Plk18) \cite{Planck:2018vyg}, and supernovae samples from either Pantheon+ \cite{Brout:2022vxf} (hereafter Pa22) or Union 3 \cite{Rubin:2023ovl} and DESY5 \cite{DES:2024jxu} releases, and the latest releases of baryonic acoustic oscillations (BAO) signature in galaxy clustering from the year one Stage IV DESI mission \cite{DESI:2024mwx} (hereafter DESI24). Within $w_0w_a$CDM as well, there were no significant change to the situation with respect to the $H_0$ and $\sigma_8$ discrepancies. Returning to $\Lambda$CDM, Sakr \cite{Sakr:2023hrl} (2022) (hereafter ZS22) investigated the hypothesis of the presence of another discrepancy, related to a difference in the matter density inferred from a local sample of cluster of galaxies surveys in a cosmological weakly dependent manner following the Oort (1958) \cite{Oort:1958nna} method, with the one obtained from CMB angular power spectrum when considering $\Lambda$CDM. They found that a low matter density is still allowed from a combination of local probes that were, either having their measurements independent of the $H_0$ and $\sigma_8$ values or made so by relaxing their related calibration parameters.  They also found that a low value for $\Omega_{\rm{M}}$ is compatible with the local values for $H_0$ and $\sigma_8$ subject of discrepancy, except when we additionally combine with the (low+high) resdshift sources composite sample Pa22. In addition, other studies at the same time or subsequent to ZS22 also hinted to such behaviors (Wagner 2022, Heisenberg 2022, Blanchard 2022, Colgain 2024,  Pedrotti 2024, Pourojaghi 2025)\cite{Wagner:2022etu, Heisenberg:2022gqk, Colgain:2024mtg, Blanchard:2022xkk, Pedrotti:2024kpn, Pourojaghi:2025vmj}, or Arico et al. (2023) \cite{Arico:2023ocu} founding a low value for $\Omega_{\rm{M}}$ from the Dark Energy Survey cosmic shear measurements when including the entire range of scales and marginalising over the baryonic parameters or more recently when using a combination of similar probes (Garcia-Garcia 2024) \cite{Garcia-Garcia:2024gzy}. Here we update on the previous study in the light of the recent release of the BAO sample by DESI24, by including this sample in our combination, following two consideration, where in the first the sound drag $r_s$ is derived within the cosmological model adopted, while we relax and consider it as a free parameter in the second variant, in order to make it more model independent in similar to what we followed in the past for the other probes we used. We also, within the two variant treatments of $r_s$, consider either the $\Lambda$CDM model (as already done in ZS22 but without DESI or other BAO samples), or the $w_0w_a$CDM model. The paper is organized as follow: in Sect.~\ref{sec:obsmethdata} we review the methods, observables and datasets used, then we show and discuss the results in Sect.~\ref{sec:resdisc} before we conclude in Sect.~\ref{sec:conclusion}.

\section{Observables, methods and datasets}\label{sec:obsmethdata}

In ZS22 (see for more details), similar to the approach followed here, we wanted to combine probes in a data driven model independent approach to avoid biases from the probes for which $H_0$ and $\sigma_8$ are showing tensions. For that, we considered, either measurements that are not or weakly derived through a cosmological model, or made so if not possible by relaxing their calibration or systematic nuisance parameters.  This was the case for the cluster counts for which we left the mass observable calibration parameter as free relaxing by then their constraints on $\Omega_{\rm{M}}$ and $\sigma_8$; or for the luminosity distance from supernovae where we also let free the light curve calibration parameter. Therefore, we did not consider the direct local measurements on $H_0$ from Cepheid stars, nor measurements on $\Omega_{\rm{M}}$ and $\sigma_8$ from CMB angular power spectrum, but instead adopted a Gaussian prior on the power spectrum amplitude parameter $A_{\rm s}$ and the spectral index $n_{\rm s}$ from Planck \cite{Planck:2018vyg} (2018) (Plk18) CMB data and a similar prior on the baryon density $\omega _{\text{b,}0}=0.0226\pm 0.00034$,  obtained by Cooke \cite{Cooke:2016rky} (2016) from Big Bang nucleosynthesis (BBN) probes. We also considered, as one of the geometric probes, the $H(z)$ measurements, introduced by Moresco \cite{moresco:2012by} (2012), that depends on the derivative of redshift with respect to cosmic time, known as cosmic chronometers (CC), and are very weakly dependent if any on the cosmological model. Practically, we use the compilation of CC data points collected from Magana et al. \cite{Magana:2017usz} and Geng, et al. \cite{Geng:2018pxk} while removing other model dependent measurements of $H(z)$ obtained from e.g. BAO observations. We shall later complement with Union II \cite{2010ApJ...716..712A} supernovae samples collected mostly at low redshifts, or a more recent collection spanning from low to high redshifts $z \in [0.01, 2.3]$, known as Pantheon+ ~\cite{Brout:2022vxf} after leaving its distance modulus calibration parameter $M_{\rm B}$ free to vary. For the growth of structure sector side, we use the cluster counts probe from a Sunayev-Zeldovich (SZ) detected clusters sample \cite{2016A&A...594A..24P} after relaxing the $(1-b)$ parameter playing the role of the calibration parameter obtained from comparison with hydrodynamical simulations. We complement by the gas mass fraction probe which corresponds to observations from massive and dynamically relaxed galaxy clusters in redshift range $0.078 \leq z \leq 1.063$ obtained by Mantz (2014) \cite{Mantz:2014xba} and for which we relax the parameter $K(z)$ as free since it is the one degenerate with the $\sigma_8$ tension. Finally, to close our 'system' of constraints, we considered direct measurements of matter density that use the mass to light ratio for galaxies in clusters divided by that of galaxies in the field as a direct proxy to $\Omega_{\rm{M}}$. In ZS22 we did not mainly include the baryonic acoustic oscillations (BAO) signature in galaxy clustering as it is not a tension free probe. Here however, we tried to relax this dependence by promoting the sound drag term to a free parameter. This allowed us to include recent BAO observations from DESI24 where we compiled  12 measurements that include the BGS sample in the redshift range $0.1 < z < 0.4$, LRG samples in $0.4 < z < 0.6$ and $0.6 < z < 0.8$, combined LRG and ELG sample in $0.8 < z < 1.1$, ELG sample in $1.1 < z < 1.6$, quasar sample in $0.8 < z < 2.1$ and the Lyman $\alpha$ Forest Sample in $1.77 < z < 4.16$.

To run Monte Carlo chains and estimate the bounds on the parameters, we use Montepython \cite{Brinckmann:2018cvx} as our inference and cosmological code, in which we implemented or used the different likelihoods. In ZS22 we limit our study to $\Lambda$CDM framework. Here we further consider the possibility of a dynamical dark energy following the Chevalier-Polarski-Linder (CPL) \cite{Linder:2002et,Chevallier:2000qy} parametrisation.
We shall show and divide our final plots, all having in common SZ clusters + $f_{\rm gas}$ clusters + $\Omega_{\rm b}$ prior from BBN + $n_{\rm s}, A_{\rm s}$ priors from CMB as our core combination of probes, in four cases based on the model we consider, whether $w_0 w_a$CDM or $\Lambda$CDM, (though the latter was already investigated in ZS22 without however additionally including DESI sample) crisscrossed with the two options, i.e. whether the $r_s$ is left free or obtained assuming one of the two models. In each case, we also subdivide in subcases based on whether we additionally include or not the Cosmic Chronometers or SN Union II sample or Pantheon+ sample or BAO measurements.

\section{Results and discussion}\label{sec:resdisc}

\subsection{within $\Lambda$CDM}

\begin{figure}[ht!]
{\psfig{file=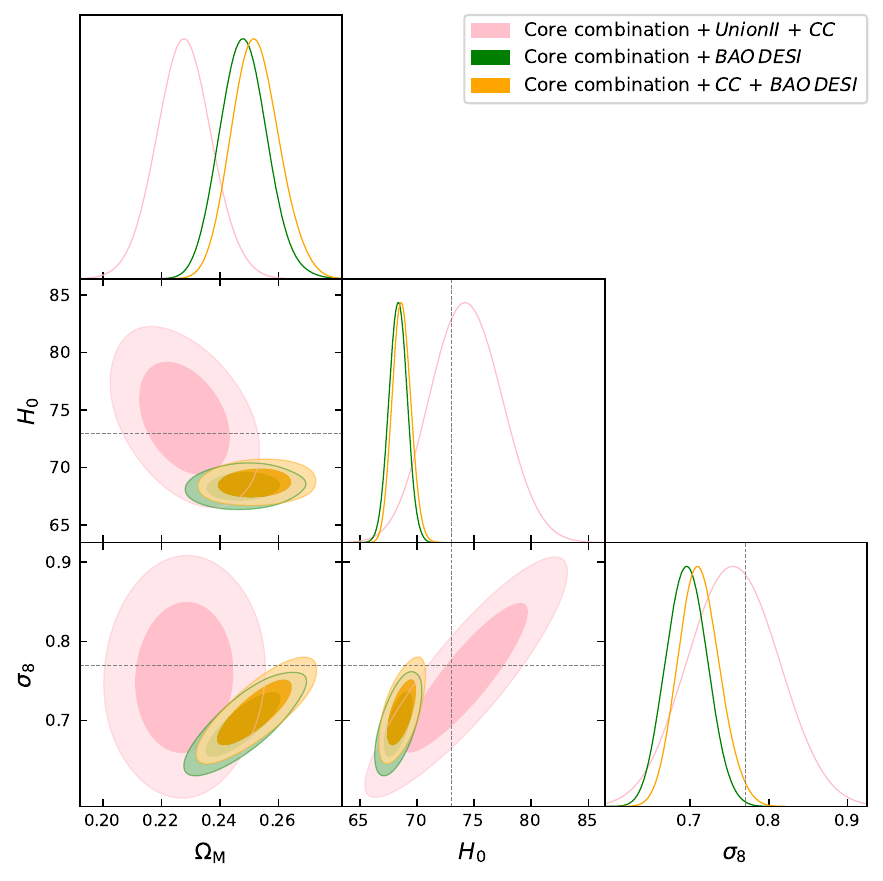,width=0.48\textwidth}}
{\psfig{file=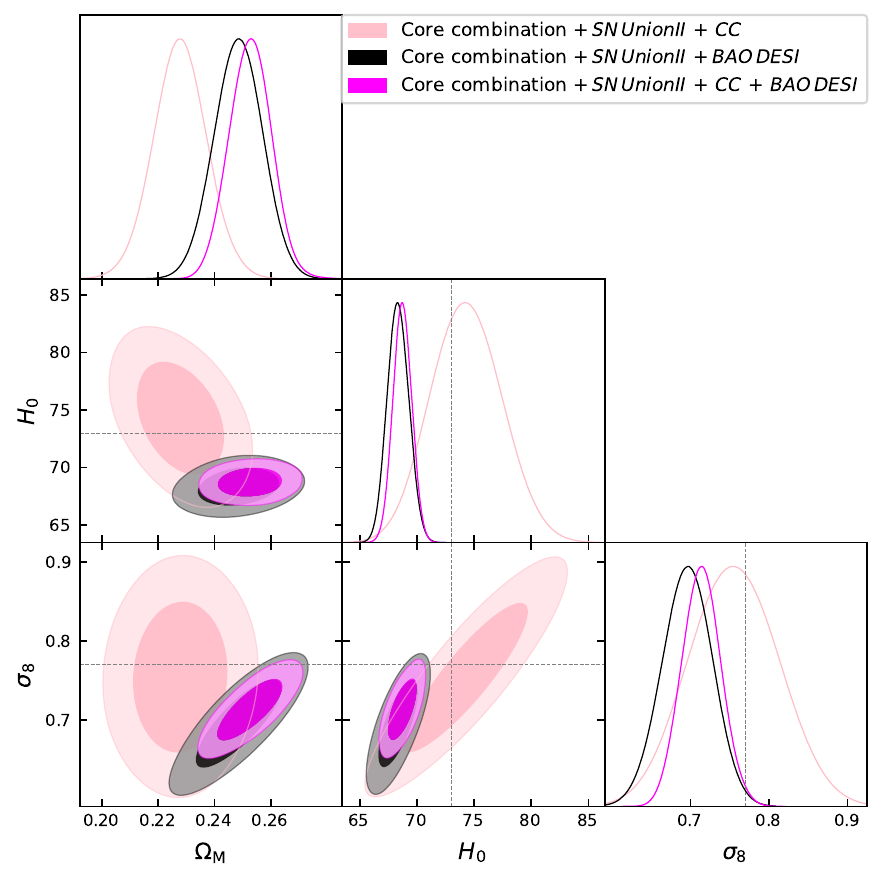,width=0.48\textwidth}}
\centerline{\psfig{file=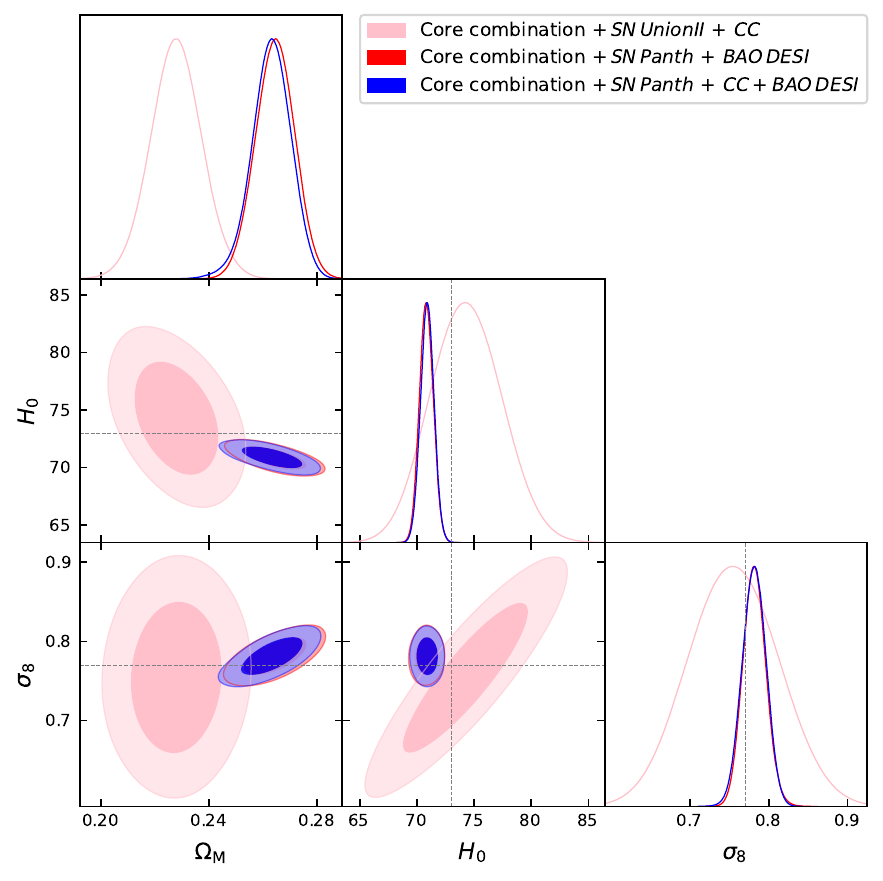,width=0.5\textwidth}}
\vspace*{8pt}
\caption{The 1D and 2D 68\% and 95\% confidence contours marginalized likelihood for the matter density $\Omega_{\rm{M}}$, the Hubble constant $H_0$ and the matter fluctuation parameter $\sigma_8$ inferred from a combinations of cluster counts and gas fraction in galaxy clusters, cosmic chronometers and astrophysical constraints on $\Omega_{\rm{M}}$ with priors from BBN measurements as well as from CMB correlations on $n_{\rm s}$ and $A_{\rm s}$. The dashed lines corresponds to $H_0$ from local observations and $\sigma_8$ from weak lensing correlations and cluster counts when fixing their calibration using hydrodynamical simulations.}
\label{fig:Om0rsfixsplit} 
\end{figure}

We start by showing in Fig.~\ref{fig:Om0rsfixsplit} an update to the compilation from ZS22 study but by further adding to all the combinations of our probes, the DESI sample, even if it is not conform to the above criteria, because we want to compare later with a case where we make it so. To better compare with ZS22, we show (pink lines) one case without DESI added where the tensions $H_0$ and $\sigma_8$ seems alleviated.  When we combine with DESI measurements, we observe in almost all cases a push to $\Omega_{\rm M}$ high but still far from CMB values. In particular, we see that no combination is able to prefer an $H_0$ and $\sigma_8$ that are both close to the ones obtained locally since combining with DESI rather show preference for $H_0$ from Plk18 although it exacerbate $\sigma_8$ tension. However, when we combine with the Pantheon+ sample, despite the fact that $\Omega_{\rm M}$ is pushed to higher values but as we said earlier, the value is still far from the one favored by CMB.

\begin{figure}[ht!]
{\psfig{file=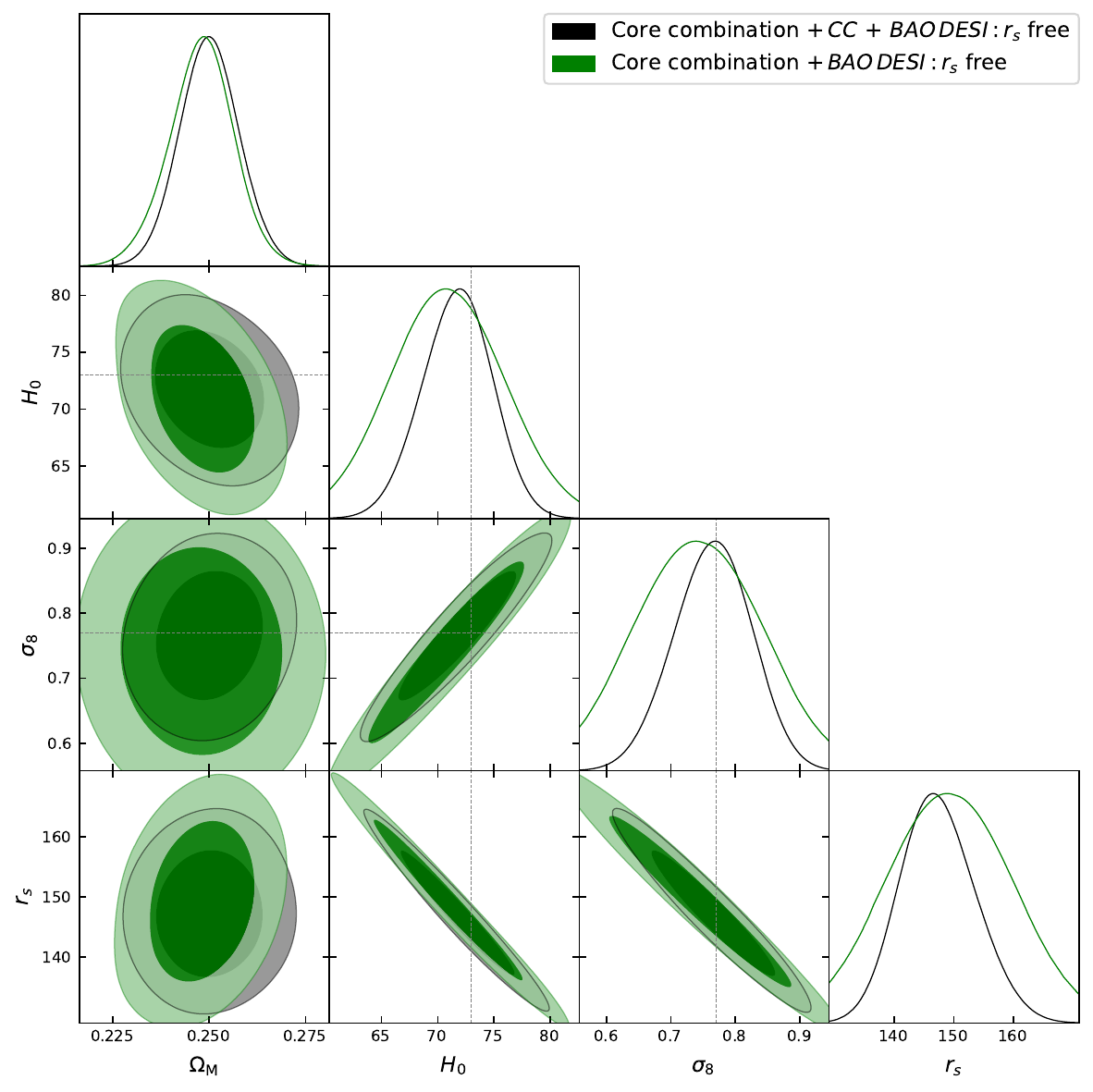,width=0.48\textwidth}}
{\psfig{file=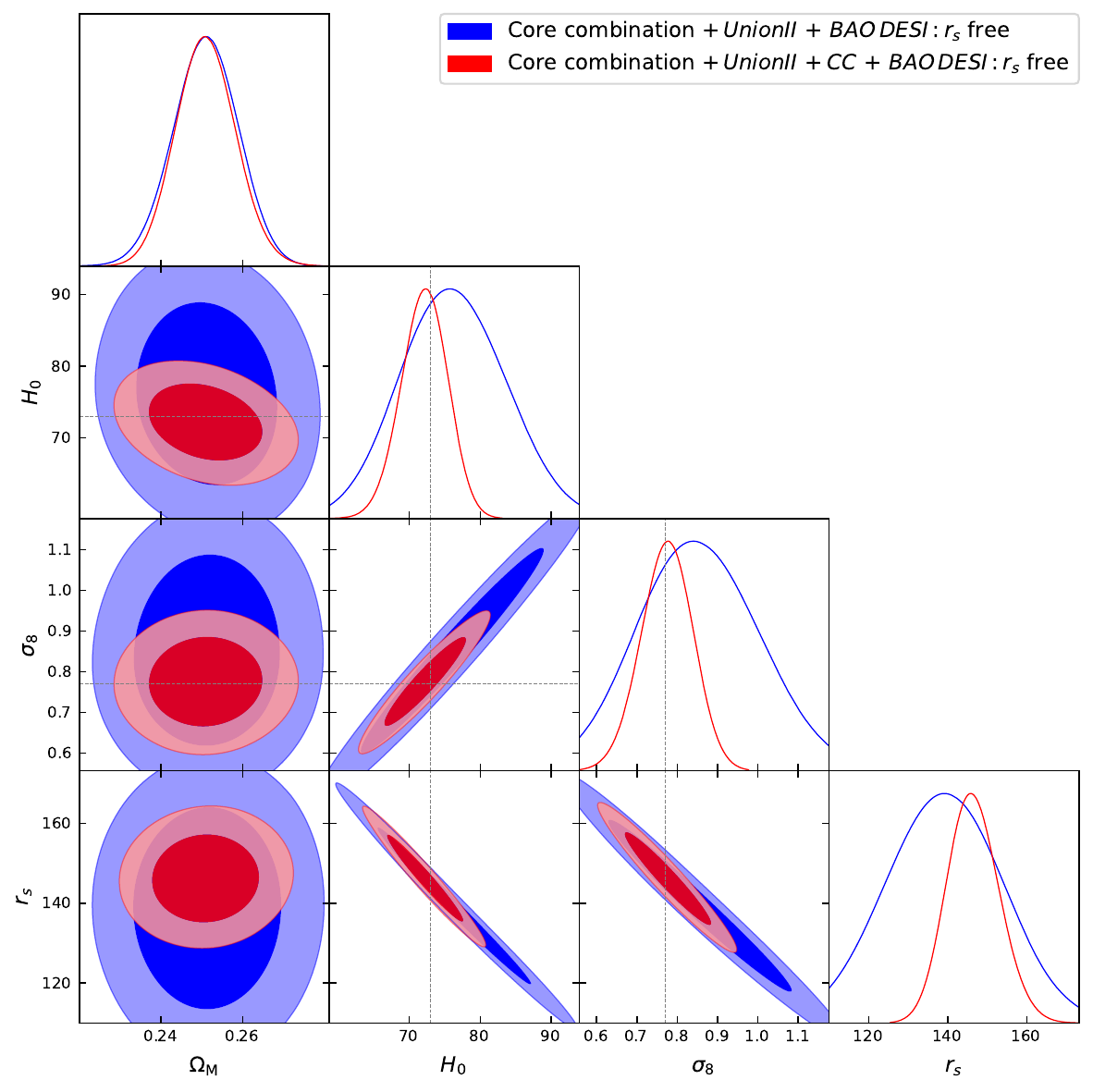,width=0.48\textwidth}}
\centerline{\psfig{file=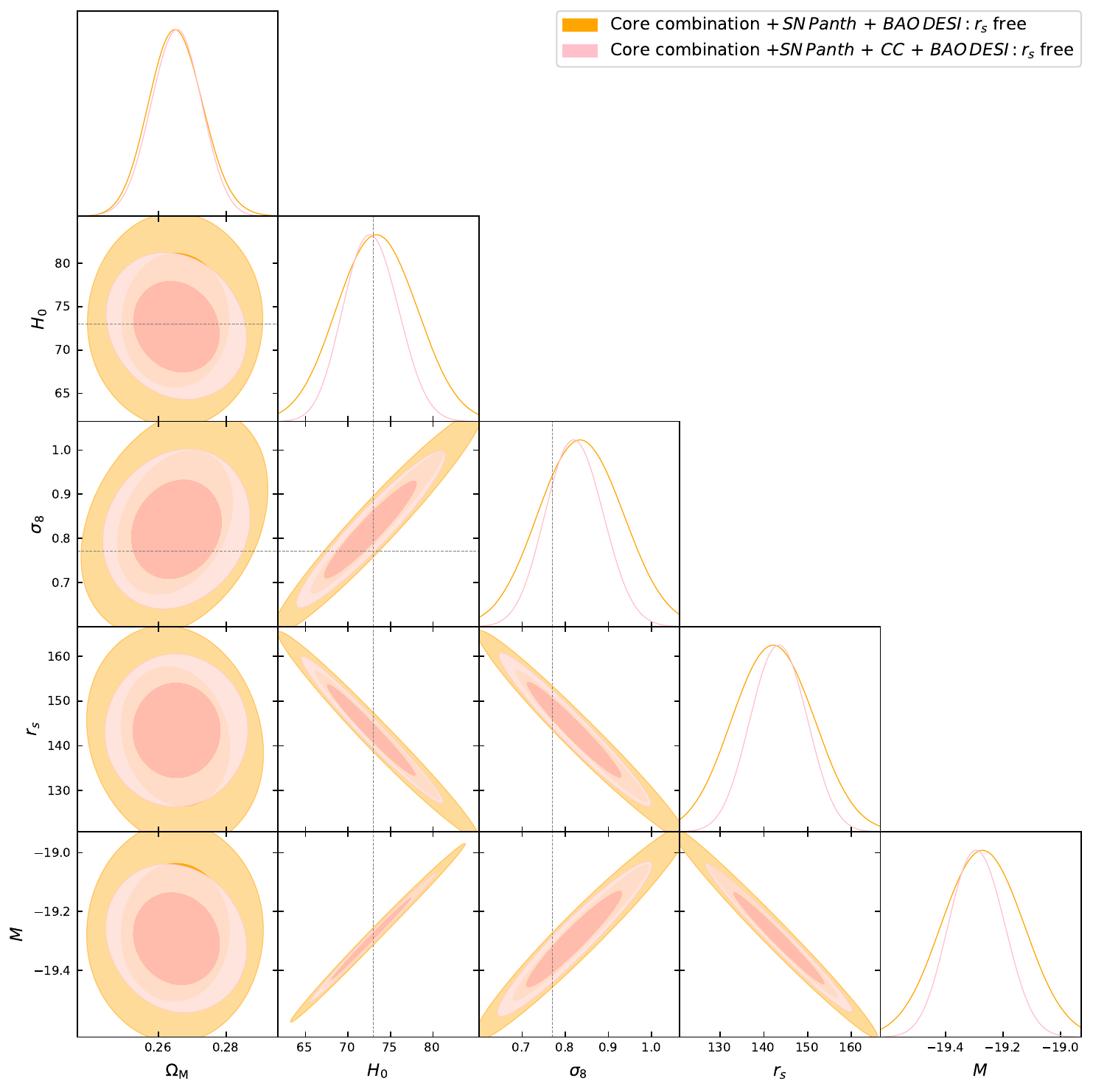,width=0.5\textwidth}}
\vspace*{8pt}
\caption{The 1D and 2D 68\% and 95\% confidence contours marginalized likelihood for the matter density $\Omega_{\rm{M}}$, the Hubble constant $H_0$ and the matter fluctuation parameter $\sigma_8$ inferred from a combinations of cluster counts and gas fraction in galaxy clusters, cosmic chronometers and astrophysical constraints on $\Omega_{\rm{M}}$ with priors from BBN measurements as well as from CMB correlations on $n_{\rm s}$ and $A_{\rm s}$.  In addition to the aforementioned probes, in the first panel up to the left, we combine with BAO datasets from DESI survey leaving the sound drag parameter $r_s$ as free, while we further add the Union II supernovae sample in the second panel up to the right, or the Pantheon+ supernovae sample in the last third panel in the bottom centre. The dashed lines corresponds to $H_0$ from local observations and $\sigma_8$ from weak lensing correlations and cluster counts when fixing their calibration using hydrodynamical simulations.}\label{fig:Om0rsfreesplit}
\end{figure}

In the case where we relax the sound drag constraints by promoting $r_s$ to become a free parameter as in Fig.~\ref{fig:Om0rsfreesplit}, we observe, when only adding DESI measurements as in the first panel without further supernovae constraints, that we gain back evidence supporting the case of a low $\Omega_{\rm{M}}$ with values of $H_0$ and $\sigma_8$ fully compatible with those from local observations. This is also the case when we add Union II supernovae sample, especially when the CC probe is included, and it remains so even when adding Pantheon+ though  $\Omega_{\rm{M}}$ is pushed to higher values. This was not the case in ZS22 where adding Pantheon+ was mitigating our hypothesis, while here the addition of DESI sample, without or especially while making $r_s$ as free, cure the problem because it balances $H_0$ and $\sigma_8$ deviation from the local values.

\subsection{Considering dynamical dark energy}

\begin{figure}[ht!]
\centerline{\psfig{file=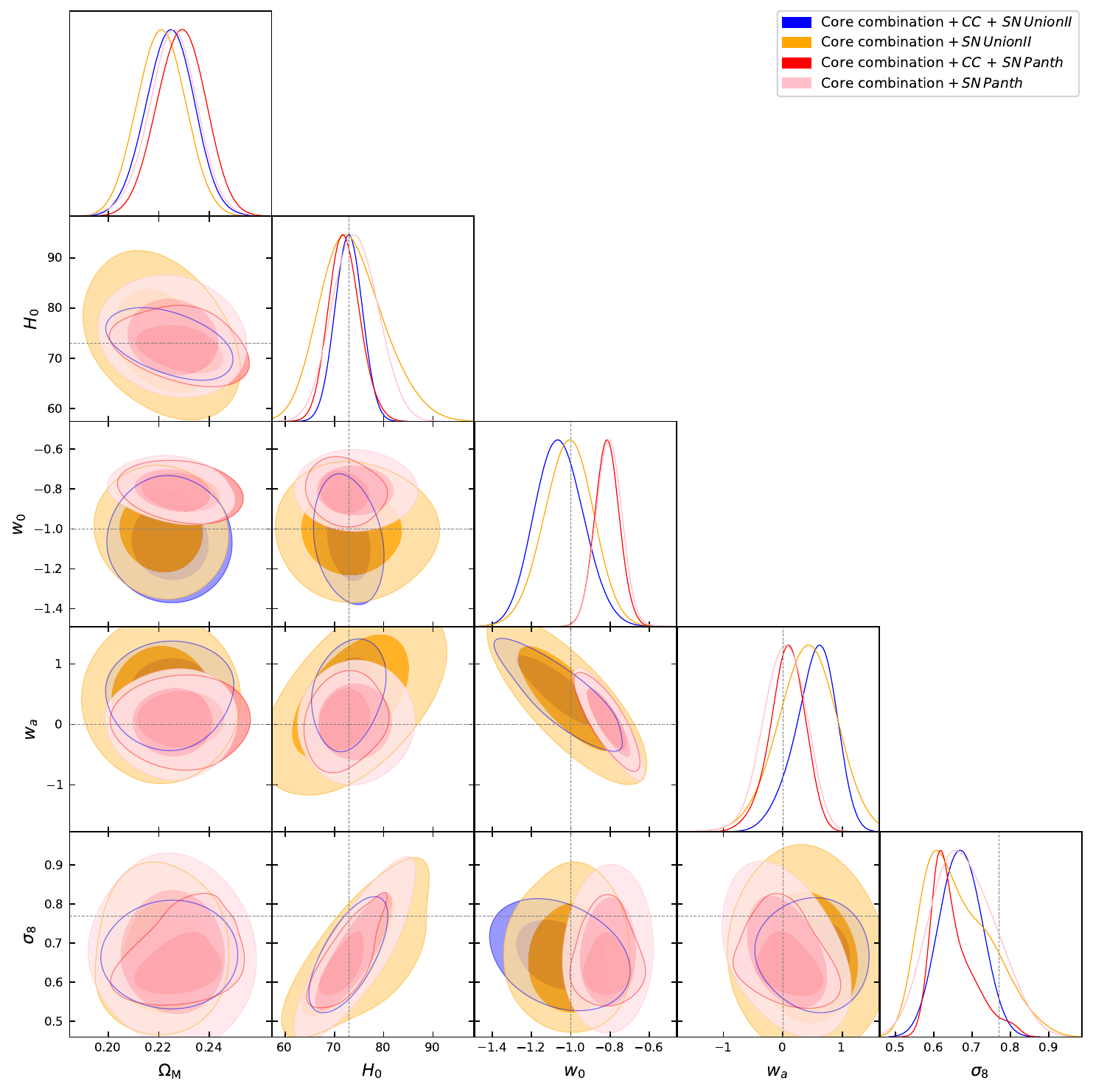,width=0.8\textwidth}}
\vspace*{8pt}
\caption{The 1D and 2D 68\% and 95\% confidence contours marginalized likelihood for the matter density $\Omega_{\rm{M}}$, the Hubble constant $H_0$, the matter fluctuation parameter $\sigma_8$ and the dark energy equation of state parameters $w_0$ and $w_a$ inferred from a combinations of cluster counts and gas fraction in galaxy clusters, cosmic chronometers and astrophysical constraints on $\Omega_{\rm{M}}$ with priors from BBN measurements as well as from CMB correlations on $n_{\rm s}$ and $A_{\rm s}$.  In addition to the aforementioned probes, we also combine with the Union II supernovae sample or the Pantheon+ supernovae sample as shown in the legend. The dashed lines corresponds to $H_0$ from local observations and $\sigma_8$ from weak lensing correlations and cluster counts when fixing their calibration using hydrodynamical simulations.}
\label{fig:Om0noDESIw0wa}
\end{figure}

In the case of a dynamical dark energy model, we start by showing in Fig.~\ref{fig:Om0noDESIw0wa} results using our core group, where we consider the CPL model with $w_0$ and $w_a$ as free parameters, before combining with DESI measurements but after adding the supernovae samples. The reason is that we want for more clarity and use of contrast to show in subsequent figures the effect of relaxing $r_s$ on the results since we saw its strong impact in the $\Lambda$CDM model assumption. As so, we observe in Fig.~\ref{fig:Om0noDESIw0wa} for all cases, whether the additional sample is Union II or Pantheon, that the $\Omega_{\rm{M}}$ is still low far from Plk18 values. $H_0$ is compatible with local values but  $\sigma_8$ is preferring very low values disfavoring our strong hypothesis. With respect to evidence for dynamical dark energy parameters far from their  $\Lambda$CDM values, we see a difference in the outcome, such that if we add Union II sample, the contours are compatible with  $\Lambda$CDM, while with Pantheon+ there is a 2$\sigma$ preference for dynamical dark energy notably for $w_0$ parameter. This difference comes from the fact that Pantheon+ includes much higher redhsift measurements and a low $\Omega_{\rm{M}}$ might be only preferred by the more close and local universe, especially that it has many supernovae in common with Union II sample at low redshifts. One thing to note for $H_0$ constraints, is that it prefers low value $\sim 0.67$ when combining CC and any of the supernovae sample along with BAO from DESI with derived $r_s$ as in the first two panels of Fig.~\ref{fig:Om0rsfixsplit} while we find again concordance with local probes when we treat $r_s$ as free parameter. 

\begin{figure}[ht!]
{\psfig{file=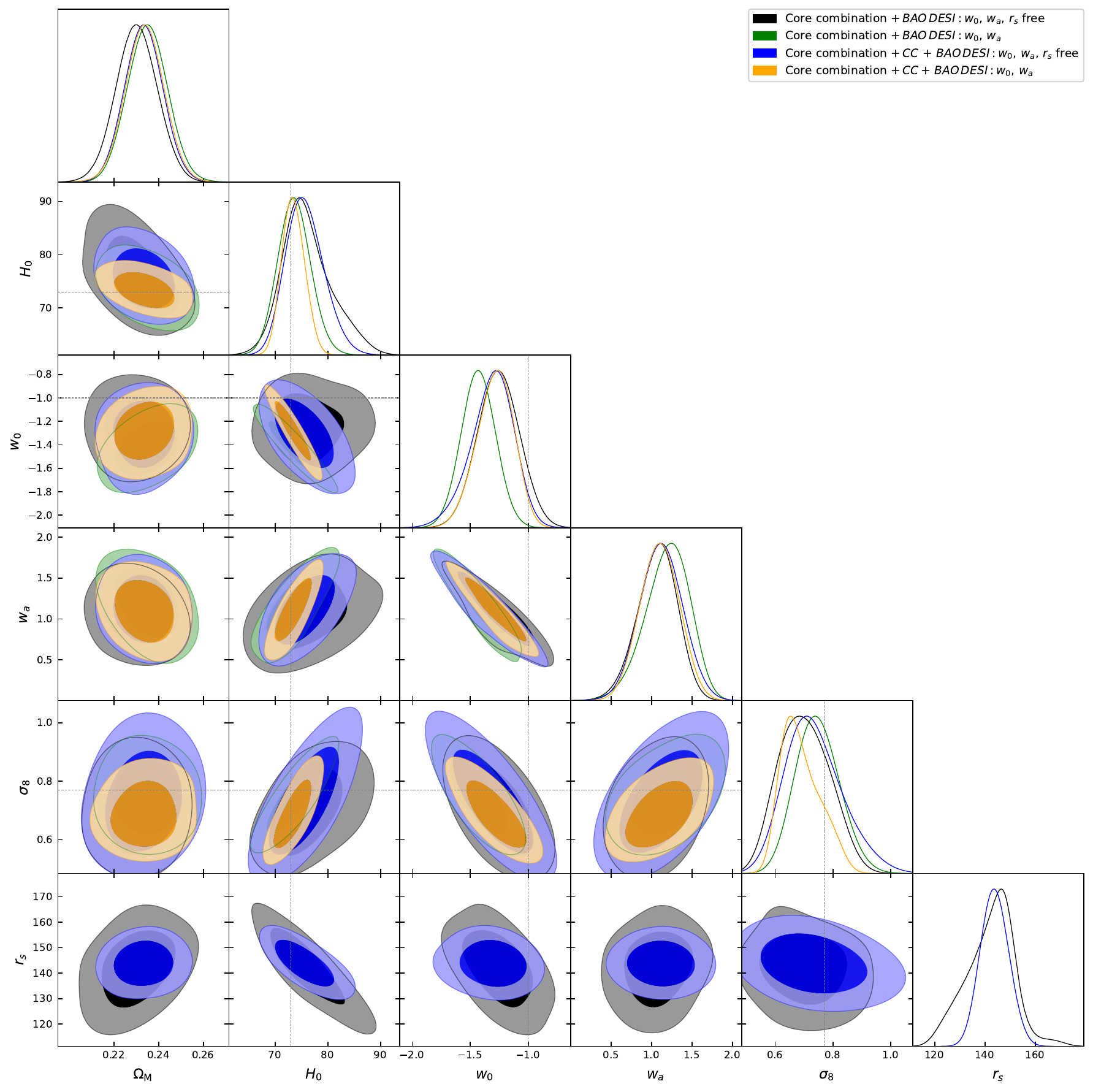,width=0.85\textwidth}}
\vspace*{8pt}
\caption{The 1D and 2D 68\% and 95\% confidence contours marginalized likelihood for the matter density $\Omega_{\rm{M}}$, the Hubble constant $H_0$, the matter fluctuation parameter $\sigma_8$ and the dark energy equation of state parameters $w_0$ and $w_a$ inferred from a combinations of cluster counts and gas fraction in galaxy clusters, cosmic chronometers and astrophysical constraints on $\Omega_{\rm{M}}$ with priors from BBN measurements as well as from CMB correlations on $n_{\rm s}$ and $A_{\rm s}$.  In addition to the aforementioned probes,  we also combine with BAO datasets from DESI survey while considering the sound drag quantity $r_s$ as free parameter. The dashed lines corresponds to $H_0$ from local observations and $\sigma_8$ from weak lensing correlations and cluster counts when fixing their calibration using hydrodynamical simulations.}\label{fig:Om0chronow0waw0wars}
\end{figure}

In the case where we add DESI sample, and compare cases where we relax or not the sound horizon constraints, we see first in Fig.~\ref{fig:Om0chronow0waw0wars} a preference for high $H_0$ and for dynamical dark energy far from $\Lambda$CDM. In all cases there is always a preference for dynamical dark energy , but $w_a$ prefers positive values while $w_0$ reaches for values $<-1$ unless when combining with SN Pantheon+ where it prefers values $>-1$.

\begin{figure}[ht!]
{\psfig{file=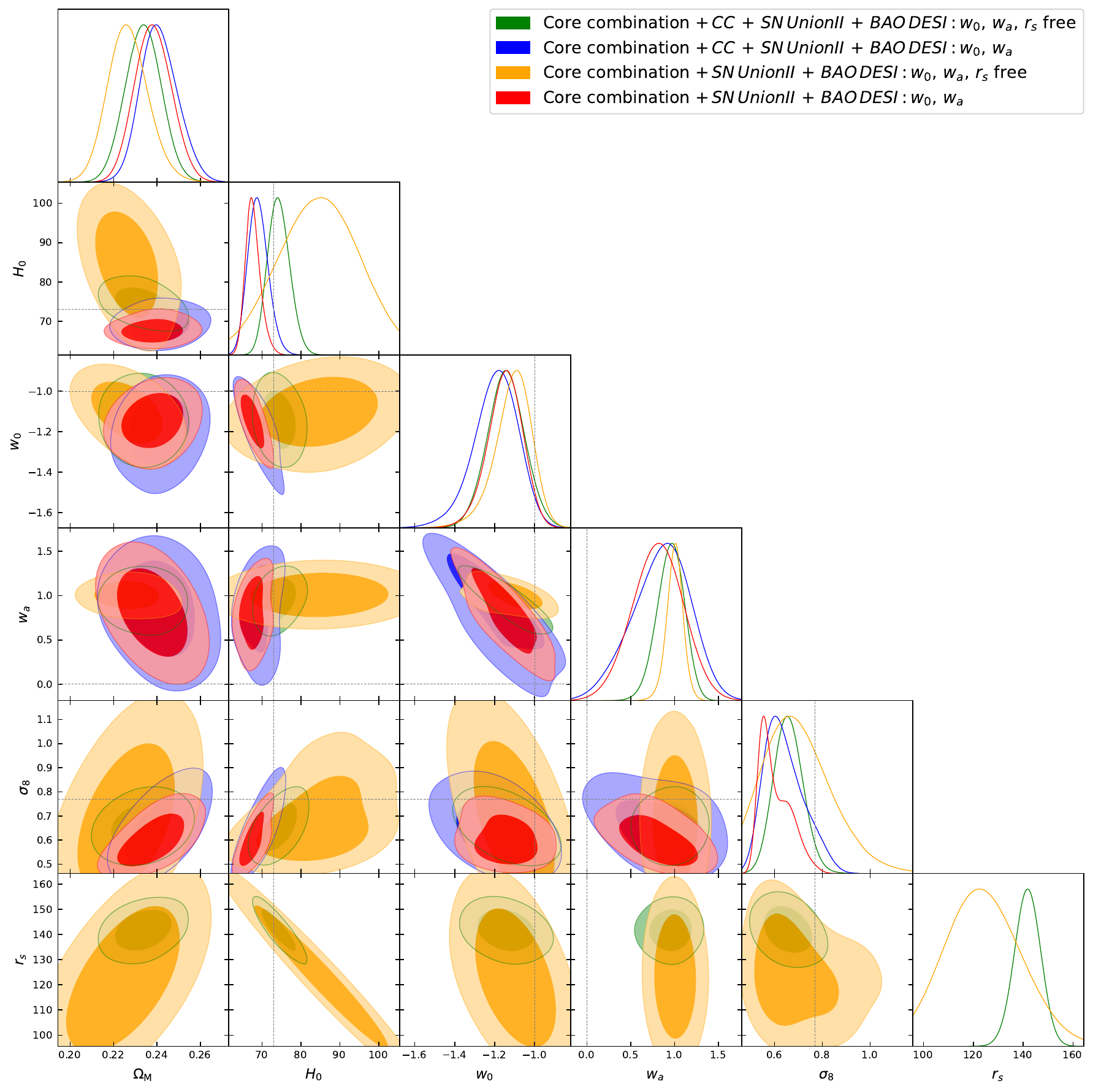,width=0.85\textwidth}}
\vspace*{8pt}
\caption{The 1D and 2D 68\% and 95\% confidence contours marginalized likelihood for the matter density $\Omega_{\rm{M}}$, the Hubble constant $H_0$, the matter fluctuation parameter $\sigma_8$ and the dark energy equation of state parameters $w_0$ and $w_a$ inferred from a combinations of cluster counts and gas fraction in galaxy clusters, cosmic chronometers and astrophysical constraints on $\Omega_{\rm{M}}$ with priors from BBN measurements as well as from CMB correlations on $n_{\rm s}$ and $A_{\rm s}$.  In addition to the aforementioned probes,  we also combine with BAO datasets from DESI survey (with or without considering the sound drag quantity $r_s$ as free parameter), and the supernovae Union II sample. The dashed lines corresponds to $H_0$ from local observations and $\sigma_8$ from weak lensing correlations and cluster counts when fixing their calibration using hydrodynamical simulations.}\label{fig:Om0Unionw0waw0wars}
\end{figure}

In the case where we add Union SN sample to the previous scenarios as in Fig.~\ref{fig:Om0Unionw0waw0wars}, we see that $H_0$ is still within its local values but $\sigma_8$ starts to become discrepant with its usual local values although it has been shifted to low values. In all cases, it is the situation, especially when we further add the CC probes, due to the fact a  low values of $\sigma_8$ has to compensate the usual low value preferred by CC for $H_0$. Only when relaxing $r_s$ and not including CC (the yellow lines), we manage to still find concordance for $H_0$ and $\sigma_8$ with other local probes. 

\begin{figure}[ht!]
{\psfig{file=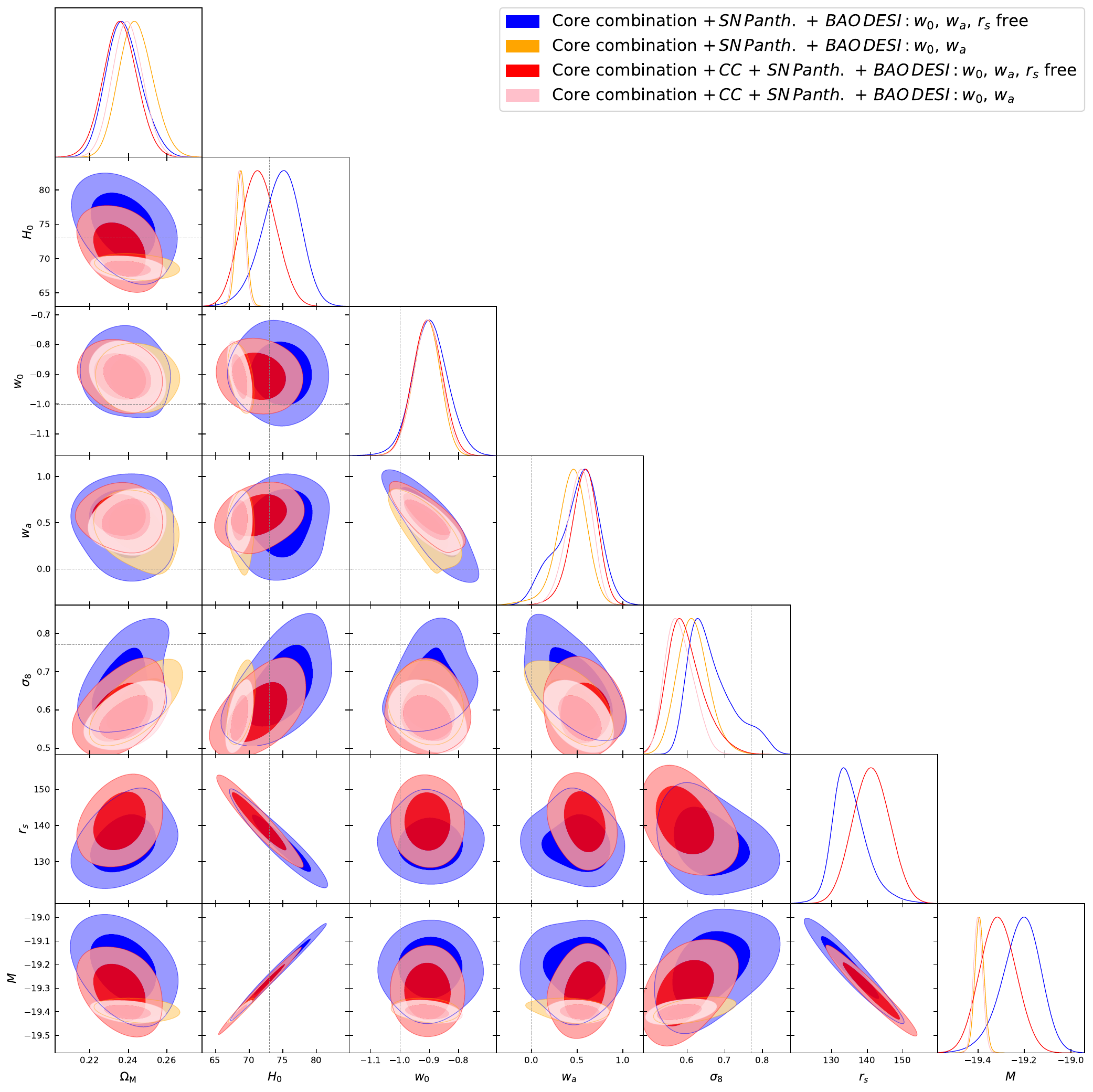,width=0.85\textwidth}}
\vspace*{8pt}
\caption{The 1D and 2D 68\% and 95\% confidence contours marginalised likelihood for the matter density $\Omega_{\rm{M}}$, the Hubble constant $H_0$, the matter fluctuation parameter $\sigma_8$ and the dark energy equation of state parameters $w_0$ and $w_a$ inferred from a combinations of cluster counts and gas fraction in galaxy clusters, cosmic chronometers and astrophysical constraints on $\Omega_{\rm{M}}$ with priors from BBN measurements as well as from CMB correlations on $n_{\rm s}$ and $A_{\rm s}$.  In addition to the aforementioned probes,  we also combine with BAO datasets from DESI survey (with or without considering the sound drag quantity $r_s$ as free parameter), and the supernovae Pantheon+ sample. The dashed lines corresponds to $H_0$ from local observations and $\sigma_8$ from weak lensing correlations and cluster counts when fixing their calibration using hydrodynamical simulations.}\label{fig:Om0Panthw0waw0wars}
\end{figure}

In the case where we rather add Pantheon+ SN sample to the previous scenarios, we observe in Fig.~\ref{fig:Om0Panthw0waw0wars} that $\sigma_8$ is more shifted to low values, however much lower than the corresponding local value as it seems that the SN sample restrict the matter fluctuation from reaching high values though we still see the same correlation orientation and trends between $\sigma_8$ and the other parameters. This also more confirmed when we add the CC probes since they usually prefer a low $H_0$ pushing as well $\sigma_8$ to lower values to compensate the high values of $H_0$ preferred before adding them.  We also note that $w_0$ $w_a$ also prefers values far from $\Lambda$CDM albeit $w_0$ goes to $<-1$ while $w_a$ stays at positive values. All the above is confirming however the tendency to loose concordance when we add probes with sources that start to reach higher redshift values such as the Pantheon+ or the BAO from DESI. \\
 
\section{Conclusion}\label{sec:conclusion}

In this work we wanted to test, in an agnostic way, the ability of a local low matter density $\Omega_{\rm M}$, in discrepancy with the value usually inferred from the CMB angular power spectrum, to accommodate observations from local probes such as not being in tension with the local values of the Hubble constant $H_0$ or the matter fluctuation $\sigma_8$ parameters. For that, we performed a Bayesian study using multiple local probes, with the criteria that they either can constrain the matter density parameter independently from the CMB constraints, or can help in doing so after making their observations more model independent by relaxing those that could be degenerate with the parameters subject of tension such as the mass observable calibration parameter for cluster counts, or by only considering Gaussian priors on parameters they directly measure such as the spectral index from CMB power spectrum, or simply because they are weakly dependent of the parameters subject of discrepancy such as the cosmic chronometers, all in the final aim to break degeneracies and auto calibrate the free non informative nuisance parameters as well as the ones subject of discrepancies. We assume however, either a dynamical dark energy model, or the standard $\Lambda$CDM  model, when computing the corresponding theoretical observables. We also include, in almost all of our Monte Carlo runs, the latest BAO measurements from the DESI year one release in addition to our core group, consisting of local matter density measurements obtained by comparing mass to light ratio in clusters over that from galaxies in the background field, and big bang nucleosynthesis prior on the baryon density, and SZ detected galaxy cluster counts, and baryon fraction in X-ray detected clusters probe. We also include in some cases, the cosmic chronometers probe or the luminosity distance measurements from supernovae Union or Pantheon+ sample.

We found that, within $\Lambda$CDM model, for different combinations of our probes, we can accommodate a low matter density along with $H_0$ and $\sigma_8$ values usually obtained from local probes, providing we promote the sound drag $r_s$ component in BAO calculations to a free parameter, and that even if we combine with the Pantheon+ supernova sample, an addition that was found in previous work to mitigate our concordance. Assuming $w_0w_a$CDM, we also found that relaxing $r_s$ allow us to accommodate $\Omega_{\rm M}$, $H_0$ and $\sigma_8$ within their local values, with still preference for $w_0w_a$ values far from $\Lambda$CDM. However, when including supernova sample, we found that the latter preference for high matter density pushes $\sigma_8$ to values much smaller than the ones usually obtained from local probes, mitigating by then a low matter density solution to the two common tensions. 

We conclude that a low matter density value, helps in preserving the concordance within $\Lambda$CDM model, even with recent BAO DESI measurements or high redshift supernovae sample, while a dynamical dark energy model ultimately fails to solve all tensions at once within our hypothesis.  Our results could be pointing to calibration issues between different probes or the way measurements are performed or the assumptions used when extracting data, on the other hand this could be indicating that $\Lambda$CDM model is facing further troubles in accommodating all the existing probes at once, while the situation is even more exacerbated if we assume $w_0w_a$CDM model and that despite that it was fully allowed within our settings to explore values for the equation of state parameters far from their $\Lambda$CDM  limit. 

\bibliographystyle{ws-ijmpd}
\bibliography{sample}

\begin{thebibliography}{10}

\bibitem{Sakr:2023hrl}
Z.~Sakr, {\em Phys. Rev. D} {\bf 108}  (2023)   083519,
  {{\ttfamily arXiv:2305.02846
  [astro-ph.CO]}}.

\bibitem{Jimenez:2024lmm}
J.~B. Jim\'enez, D.~Bettoni, D.~Figueruelo and F.~A. Teppa~Pannia, {\em Phys.
  Dark Univ.} {\bf 47}  (2025)   101761,
  {{\ttfamily arXiv:2410.18645
  [astro-ph.CO]}}.

\bibitem{Brito:2024bhh}
L.~S. Brito, J.~F. Jesus, A.~A. Escobal and S.~H. Pereira (12 2024)
  {{\ttfamily arXiv:2412.06756
  [astro-ph.CO]}}.

\bibitem{Tang:2024lmo}
X.~T. Tang, D.~Brout, T.~Karwal, C.~Chang, V.~Miranda and M.~Vincenzi (12 2024)
  {{\ttfamily arXiv:2412.04430
  [astro-ph.CO]}}.

\bibitem{Mirpoorian:2024fka}
S.~H. Mirpoorian, K.~Jedamzik and L.~Pogosian (11 2024)
  {{\ttfamily arXiv:2411.16678
  [astro-ph.CO]}}.

\bibitem{Pang:2024wul}
Y.-H. Pang, X.~Zhang and Q.-G. Huang (11 2024)
  {{\ttfamily arXiv:2411.14189
  [astro-ph.CO]}}.

\bibitem{Ye:2024zpk}
G.~Ye (11 2024) {{\ttfamily
  arXiv:2411.11743 [astro-ph.CO]}}.

\bibitem{Zhumabek:2024tvp}
T.~Zhumabek, A.~Mukhamediya, H.~Chakrabarty and D.~Malafarina (11 2024)
  {{\ttfamily arXiv:2411.05965
  [astro-ph.CO]}}.

\bibitem{DeSimone:2024lvy}
B.~De~Simone, M.~H. P.~M. van Putten, M.~G. Dainotti and G.~Lambiase, {\em
  JHEAp} {\bf 45}  (2025) 290,
  {{\ttfamily arXiv:2411.05744
  [astro-ph.CO]}}.

\bibitem{Simon:2024jmu}
T.~Simon, T.~Adi, J.~L. Bernal, E.~D. Kovetz, V.~Poulin and T.~L. Smith (10
  2024) {{\ttfamily arXiv:2410.21459
  [astro-ph.CO]}}.

\bibitem{Toda:2024uff}
Y.~Toda and O.~Seto (10 2024) {{\ttfamily
  arXiv:2410.21925 [astro-ph.CO]}}.

\bibitem{SolaPeracaula:2024iil}
J.~Sol\`a~Peracaula, { {Composite running vacuum in the Universe: implications
  on the cosmological tensions}}, in {\em {17th Marcel Grossmann Meeting}: {On
  Recent Developments in Theoretical and Experimental General Relativity,
  Gravitation, and Relativistic Field Theories}\/},  (10 2024).
{{\ttfamily arXiv:2410.20382
  [astro-ph.CO]}}.

\bibitem{Toda:2024ncp}
Y.~Toda, W.~Giar\`e, E.~\"Oz\"ulker, E.~Di~Valentino and S.~Vagnozzi, {\em
  Phys. Dark Univ.} {\bf 46}  (2024)   101676,
  {{\ttfamily arXiv:2407.01173
  [astro-ph.CO]}}.

\bibitem{Yadav:2024duq}
A.~Yadav, S.~Kumar, C.~Kibris and O.~Akarsu (6 2024)
  {{\ttfamily arXiv:2406.18496
  [astro-ph.CO]}}.

\bibitem{Liu:2024vlt}
Y.~Liu, H.~Yu and P.~Wu, {\em Phys. Rev. D} {\bf 110}  (2024)   L021304,
  {{\ttfamily arXiv:2406.02956
  [astro-ph.CO]}}.

\bibitem{zabat:2024wof}
S.~zabat, Y.~Kehal and K.~Nouicer (5 2024)
  {{\ttfamily arXiv:2405.20240 [gr-qc]}}.

\bibitem{HosseiniMansoori:2024pdq}
S.~A. Hosseini~Mansoori and H.~Moshafi, {\em Astrophys. J.} {\bf 975}  (2024)
  275, {{\ttfamily arXiv:2405.05843
  [astro-ph.CO]}}.

\bibitem{Sakr:2024eee}
Z.~Sakr and L.~Schey, {\em JCAP} {\bf 10}  (2024)   052,
  {{\ttfamily arXiv:2405.03627
  [astro-ph.CO]}}.

\bibitem{Gomez-Valent:2024tdb}
A.~Gomez-Valent and J.~Sol\`a~Peracaula, {\em Astrophys. J.} {\bf 975}  (2024)
  ~64, {{\ttfamily arXiv:2404.18845
  [astro-ph.CO]}}.

\bibitem{Cook:2024eff}
R.~J. Cook, {\em Astrophys. J.} {\bf 965}  (2024)   127,
  {{\ttfamily arXiv:2404.13712 [gr-qc]}}.

\bibitem{Yeung:2024krv}
S.~Yeung, W.~Zhang and M.-c. Chu (3 2024)
  {{\ttfamily arXiv:2403.11499
  [hep-ph]}}.

\bibitem{Akarsu:2024qiq}
O.~Akarsu, E.~O. Colg\'ain, A.~A. Sen and M.~M. Sheikh-Jabbari, {\em Universe}
  {\bf 10}  (2024)   305, {{\ttfamily
  arXiv:2402.04767 [astro-ph.CO]}}.

\bibitem{Khosravi:2023rhy}
N.~Khosravi, {\em Phys. Rev. D} {\bf 110}  (2024)   063507,
  {{\ttfamily arXiv:2312.13886
  [astro-ph.CO]}}.

\bibitem{Tutusaus:2023cms}
I.~Tutusaus, M.~Kunz and L.~Favre (11 2023)
  {{\ttfamily arXiv:2311.16862
  [astro-ph.CO]}}.
  
\bibitem{Sakr:2023mao}
Z.~Sakr (4 2023)
  {{\ttfamily arXiv:2305.02913
  [astro-ph.CO]}}.   
    
  
\bibitem{Yarahmadi:2024lzd}
M.~Yarahmadi, A.~Salehi {\em Eur. Phys. J. C. } (12 2024)
  {{\ttfamily arXiv:2501.07860
  [astro-ph.CO]}}.  
 
\bibitem{Shah:2024gfu}
 R~Shah {\em et al.} (12 2024)
  {{\ttfamily arXiv:2412.14750
  [astro-ph.CO]}}.

\bibitem{Barua:2024gei}
S.~Barua, S.~Desai (12 2024)
  {{\ttfamily arXiv:2412.19240
  [astro-ph.CO]}}.

\bibitem{Sadeghnezhad:2025txo}
N.~Sadeghnezhad (1 2025)
  {{\ttfamily arXiv:2501.02457
  [astro-ph.CO]}}.

\bibitem{Yarahmadi:2024afr}
M.~Yarahmadi (1 2025) {\em Phys. Dark Univ.}
  {{\ttfamily arXiv:2501.06611
  [astro-ph.CO]}}.
   
\bibitem{Planck:2018vyg}
 Planck Collaboration (N.~Aghanim {\em et~al.}), {\em Astron. Astrophys.} {\bf
  641}  (2020)  ~A6, {{\ttfamily
  arXiv:1807.06209 [astro-ph.CO]}}, [Erratum: Astron.Astrophys. 652, C4
  (2021)].

\bibitem{Brout:2022vxf}
D.~Brout {\em et~al.}, {\em Astrophys. J.} {\bf 938}  (2022)   110,
  {{\ttfamily arXiv:2202.04077
  [astro-ph.CO]}}.

\bibitem{Rubin:2023ovl}
D.~Rubin {\em et~al.} (11 2023)
  {{\ttfamily arXiv:2311.12098
  [astro-ph.CO]}}.

\bibitem{DES:2024jxu}
 DES Collaboration (T.~M.~C. Abbott {\em et~al.}), {\em Astrophys. J. Lett.}
  {\bf 973}  (2024)   L14, {{\ttfamily
  arXiv:2401.02929 [astro-ph.CO]}}.

\bibitem{DESI:2024mwx}
 DESI Collaboration (A.~G. Adame {\em et~al.}) (4 2024)
  {{\ttfamily arXiv:2404.03002
  [astro-ph.CO]}}.

\bibitem{Oort:1958nna}
J.~H. Oort, { {Distribution of galaxies and the density in the universe}}, in
  {\em {11\`eme Conseil de Physique de l'Institut International de Physique
  Solvay}: {La structure et l'\'evolution de l'univers : rapports et
  discussions}\/},  (1958), pp. 163--184.

\bibitem{Wagner:2022etu}
J.~Wagner (3 2022) {{\ttfamily
  arXiv:2203.11219 [astro-ph.CO]}}.

\bibitem{Heisenberg:2022gqk}
L.~Heisenberg, H.~Villarrubia-Rojo and J.~Zosso, {\em Phys. Rev. D} {\bf 106}
  (2022)   043503, {{\ttfamily
  arXiv:2202.01202 [astro-ph.CO]}}.

\bibitem{Colgain:2024mtg}
E.~O. Colg\'ain and M.~M. Sheikh-Jabbari (12 2024)
  {{\ttfamily arXiv:2412.12905
  [astro-ph.CO]}}.

\bibitem{Blanchard:2022xkk}
A.~Blanchard, J.-Y. H\'eloret, S.~Ili\'c, B.~Lamine and I.~Tutusaus, {\em Open
  J. Astrophys.} {\bf 7}  (2024)   117170,
  {{\ttfamily arXiv:2205.05017
  [astro-ph.CO]}}.

\bibitem{Pedrotti:2024kpn}
D.~Pedrotti, J.-Q. Jiang, L.~A. Escamilla, S.~S. da~Costa and S.~Vagnozzi, {\em
  Phys. Rev. D} {\bf 111}  (2025)   023506,
  {{\ttfamily arXiv:2408.04530
  [astro-ph.CO]}}.
  
\bibitem{Pourojaghi:2025vmj}
S.~Pourojaghi, Saeed, M.~Malekjani (1 2025)
  {{\ttfamily arXiv:2501.05740
  [astro-ph.CO]}}.
  
\bibitem{Arico:2023ocu}
G.~Aric\`o, R.~E. Angulo, M.~Zennaro, S.~Contreras, A.~Chen and
  C.~Hern\'andez-Monteagudo, {\em Astron. Astrophys.} {\bf 678}  (2023)   A109,
  {{\ttfamily arXiv:2303.05537
  [astro-ph.CO]}}.

\bibitem{Garcia-Garcia:2024gzy}
C.~Garc\'\i{}a-Garc\'\i{}a, M.~Zennaro, G.~Aric\`o, D.~Alonso and R.~E. Angulo,
  {\em JCAP} {\bf 08}  (2024)   024,
  {{\ttfamily arXiv:2403.13794
  [astro-ph.CO]}}.

\bibitem{Cooke:2016rky}
R.~J. Cooke, M.~Pettini, K.~M. Nollett and R.~Jorgenson, {\em Astrophys. J.}
  {\bf 830}  (2016)   148, {{\ttfamily
  arXiv:1607.03900 [astro-ph.CO]}}.

\bibitem{moresco:2012by}
M.~Moresco, L.~Verde, L.~Pozzetti, R.~Jimenez and A.~Cimatti, {\em JCAP} {\bf
  07}  (2012)   053, {{\ttfamily
  arXiv:1201.6658 [astro-ph.CO]}}.

\bibitem{Magana:2017usz}
J.~Magana, V.~Motta, V.~H. Cardenas and G.~Foex, {\em Mon. Not. Roy. Astron.
  Soc.} {\bf 469}  (2017) 47, {{\ttfamily
  arXiv:1703.08521 [astro-ph.CO]}}.

\bibitem{Geng:2018pxk}
J.-J. Geng, R.-Y. Guo, A.~Wang, J.-F. Zhang and X.~Zhang, {\em Commun. Theor.
  Phys.} {\bf 70}  (2018)   445,
  {{\ttfamily arXiv:1806.10735
  [astro-ph.CO]}}.

\bibitem{2010ApJ...716..712A}
R.~{Amanullah}, C.~{Lidman}, D.~{Rubin}, G.~{Aldering}, P.~{Astier},
  K.~{Barbary}, M.~S. {Burns}, A.~{Conley}, K.~S. {Dawson}, S.~E. {Deustua},
  M.~{Doi}, S.~{Fabbro}, L.~{Faccioli}, H.~K. {Fakhouri}, G.~{Folatelli}, A.~S.
  {Fruchter}, H.~{Furusawa}, G.~{Garavini}, G.~{Goldhaber}, A.~{Goobar}, D.~E.
  {Groom}, I.~{Hook}, D.~A. {Howell}, N.~{Kashikawa}, A.~G. {Kim}, R.~A.
  {Knop}, M.~{Kowalski}, E.~{Linder}, J.~{Meyers}, T.~{Morokuma}, S.~{Nobili},
  J.~{Nordin}, P.~E. {Nugent}, L.~{{\"O}stman}, R.~{Pain}, N.~{Panagia},
  S.~{Perlmutter}, J.~{Raux}, P.~{Ruiz-Lapuente}, A.~L. {Spadafora},
  M.~{Strovink}, N.~{Suzuki}, L.~{Wang}, W.~M. {Wood-Vasey}, N.~{Yasuda} and
  T.~{Supernova Cosmology Project}, {\bf 716} (June 2010) 712,
  {{\ttfamily arXiv:1004.1711
  [astro-ph.CO]}}.

\bibitem{2016A&A...594A..24P}
{Planck Collaboration}, P.~A.~R. {Ade}, N.~{Aghanim}, M.~{Arnaud},
  M.~{Ashdown}, J.~{Aumont}, C.~{Baccigalupi}, A.~J. {Banday}, R.~B.
  {Barreiro}, J.~G. {Bartlett} and et~al., {\em aap} {\bf 594} (September 2016)
    A24.

\bibitem{Mantz:2014xba}
A.~B. Mantz, S.~W. Allen, R.~G. Morris, D.~A. Rapetti, D.~E. Applegate, P.~L.
  Kelly, A.~von~der Linden and R.~W. Schmidt, {\em Mon. Not. Roy. Astron. Soc.}
  {\bf 440}  (2014) 2077, {{\ttfamily
  arXiv:1402.6212 [astro-ph.CO]}}.

\bibitem{Brinckmann:2018cvx}
T.~Brinckmann and J.~Lesgourgues, {\em Phys. Dark Univ.} {\bf 24}  (2019)
  100260, {{\ttfamily arXiv:1804.07261
  [astro-ph.CO]}}.

\bibitem{Linder:2002et}
E.~V. Linder, {\em Phys. Rev. Lett.} {\bf 90}  (2003)   091301,
  {{\ttfamily
  arXiv:astro-ph/0208512}}.

\bibitem{Chevallier:2000qy}
M.~Chevallier and D.~Polarski, {\em Int. J. Mod. Phys. D} {\bf 10}  (2001) 213,
  {{\ttfamily arXiv:gr-qc/0009008}}.

\end{thebibliography}

\end{document}